# The role of energetic flow structures on the aeolian transport of sediment and plastic debris

Manousos Valyrakis[1,2]*, Xiao-Hu Zhao [3], Thomas Pähtz[4,5], Zhen-Shan Li[6,7]

[1] Department of Civil Engineering, Aristotle University of Thessaloniki, Thessaloniki 54124, Greece.

[2] School of Engineering, University of Glasgow, Glasgow G128QQ, UK.

[3] Division of Environmental Management and Policy, School of Environment, Tsinghua University, Beijing 100084, China.

[4] Institute of Port, Coastal and Offshore Engineering, Ocean College, Zhejiang University, 316021 Zhoushan, China.

[5] Donghai Laboratory, 316021, Zhoushan, China

[6] College of Environmental Science and Engineering, Peking University, Beijing 100871, China.

[7] The Key Laboratory of Water and Sediment Sciences, Ministry of Education, Peking University, Beijing 100871, China.

*Corresponding author: Manousos Valyrakis mvalyra@civil.auth.gr

**Highlights**

- An event-based model is developed to assess incipient motion conditions of natural coarse particles.
- The incipient motion of plastic spheres is directly investigated in a wind tunnel.
- The role of turbulent fluctuations is shown for creep transport and mechanical sieving.

**Author Contributions**

Manousos Valyrakis and Xiao-Hu Zhao designed the research. Manousos Valyrakis and Xiao-Hu Zhao wrote a first draft of the manuscript. Xiao-Hu Zhao set up the experimentsand performed the measurements. Manousos Valyrakis, Xiao-Hu Zhao and Thomas Pähtz analyzed the data and discussed the results. Manousos Valyrakis, Xiao-Hu Zhao and Thomas Pähtz organized and revised later versions of the manuscript. Zhen-Shan Li and Manousos Valyrakis supervised Xiao-Hu Zhao and directed the project. Zhen-Shan Li and Thomas Pähtz acquired funding for this project.

**Graphical Abstract and Text**

Researchers studied the movement of plastic spheres in a wind tunnel to understand how wind turbulence affects their motion. They have developed a model based on energetic, turbulent events to understand how natural coarse sediment starts to move. The study found that turbulent wind gusts play a crucial role in moving these particles, either by rolling them downwind or by shorter rocking motions resulting in creeping and mechanical sorting. It emphasizes the probabilistic nature of sediment and plastic debris entrainment and contributes towards a better understanding of aeolian transport processes.

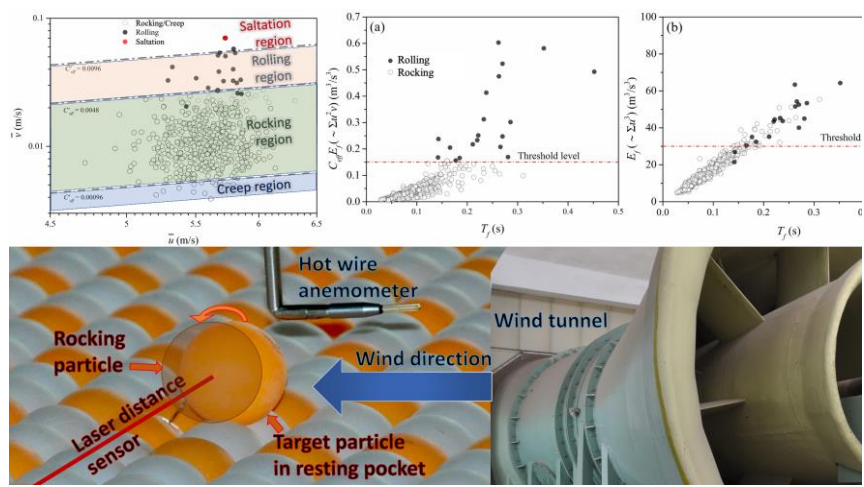

**Abstract**

Recently, significant progress has been made in conceptually describing the dynamic aspects of coarse particle entrainment, which has been explored experimentally for open channel flows. The aim of this study is to extend the application of energy criterion to the low mobility aeolian transport of solids (including both natural sediment and anthropogenic debris such as plastics), ranging from incomplete (rocking) to full (rolling) entrainments. This is achieved by linking particle movements to energetic flow events, which are defined as flow structures with the ability to work on particles, setting them into motion. It is hypothesized that such events should impart sufficient energy to the particles, above a certain threshold value. The concept's validity is demonstrated experimentally, using a wind tunnel and laser distance sensor (LDS) to capture the dynamics of an individual target particle, exposed on a rough bed surface. Measurements are acquired at a high spatiotemporal resolution, and synchronously with the instantaneous air velocity at an appropriate distance upwind of the target particle, using a hot film anemometer. This enables the association of flow events with rocking and rolling entrainments. Furthermore, it is shown that rocking and rolling may have distinct energy thresholds. Estimates of the energy transfer efficiency, normalized by the drag coefficient, range over an order of magnitude (from about 0.001 to 0.0048 for rocking, up to about 0.01, for incipient rolling). The proposed event-based theoretical framework is a novel approach to characterizing the energy imparted from the wind to the soil surface and could have potential implications for modelling intermittent creep transport of coarse particles and related aeolian bedforms.

# 1. Introduction

Transport of coarse quartz particles by different media, such as water and air, constitutes a major challenge for hydraulic engineering and earth surface dynamics. Amongst these processes, the transport of solids by air, namely aeolian transport, is responsible to a significant degree for the many metric tonnes of sediment reaching the oceans yearly. Similar to transport by water, rounded sand to gravel particles are typically carried downwind by rolling or saltation mode. Many researchers have studied aeolian transport experimentally in wind tunnels, primarily focusing on the motion of saltation sand particles [1-5].

Compared with saltation, rolling has not been adequately explored, even though it may account for up to 25% of the aeolian transport [6] and may play a significant role in the formation of aeolian megaripples [7-11], as well as for erosion control [12-14]. The entrainment of such grains, termed surface creep, may result from short-lived but energetic wind bursts or impacts with saltation particles [15-16]. Likewise, the study of rocking (or "wobbling") particle motions has typically received less attention, as such incomplete motions do not result in net transport downwind. However, a recent study [17] found the mechanical sieving of rocking gravel particles to be instrumental in the evolution of gravel-mantled megaripples of the Argentinean Puna. Further, customarily the threshold for rolling is four to six times lower than the threshold for direct entrainment into saltation, and it is commonly observed that grains may roll over their neighbours before entering the saltation regime [18, 19]. Thus, investigating the role of incomplete (rocking) and full rolling motions may find application from soil erosion control to crop technologies and is essential for gaining a holistic understanding of wind-induced geomorphologic processes.

The initiation of particle motions due to turbulent wind is represented by unsteady rocking, rolling, and take-off. Several researchers have experimentally observed the incipient motions of sediments. Bisal and Nielsen [20] used a binocular microscope to detect particles' oscillatory motion (rocking) back and forth before getting entrained. A more detailed wind tunnel study by Lyles and Krauss [21] involved counting particle minute movements using a telescope. They found that near the threshold wind speed, particle motions occurred intermittently at an average frequency of about 1.8 Hz, which they postulated to be associated with the maximum energy-containing frequencies from the turbulence spectra. Williams et al. [22] conducted similar experiments, where a rocking-rolling and take-off mode for aeolian transport of particles has been further confirmed. The above wind tunnel investigations are limited to relatively small sand grains ranging from 0.1 mm to 1.7 mm. However, the field investigation shows that gravel-size sediments up to several centimetres have similar modes of incipient motions to sand grains [17]. In addition, rocking and rolling have also been suggested as the basic modes of motion of particles ranging from fine sand to gravel on Venus [23]. Thus, it becomes apparent that the rocking and rolling motions of individual grains exposed to the wind are an important mechanism for conducting geomorphic work.

The above studies are limited to only offering a qualitative description of particle motion dynamics solely by visual observation. A more comprehensive study of a coarse particle's dynamic rocking response would involve obtaining detailed quantitative features for the full range of particle motion (rocking to rolling), such as the frequency and amplitude of its displacement.

This study aims to address the above by employing the energy criterion and appropriately designed wind tunnel experiments. An energy framework was initially established for the incipient motion of coarse particles in water [24]. Though it has also been applied to theoretically model aeolian entrainment [25-26], a direct experimental examination is still missing.

We perform experiments to measure the dynamical features of rocking particles' entrainment near incipient-motion flow conditions. This is achieved by using a high-resolution laser distance detector to monitor the time series of the angular displacement of a target particle (medium gravel-size) resting on an artificially roughened wind tunnel test section. Further, the particle's displacement measurements are recorded synchronously with the instantaneous wind velocities upwind near its vicinity, which allows for associating highly energetic events with different levels of the particle's response. Finally, analysis of the obtained data helps verify the presented theoretical energy criterion for aeolian transport.

## 2. Re-examination of aerodynamic force models for aeolian transport

It is customarily believed that aeolian transport occurs when the average aerodynamic forces overcome the resistance of surface sand grains. Based on this concept, a deterministic theoretical framework for the entrainment of loose coarse particles from a flat surface was first developed by Bagnold [6]. By relating surface shear stress to air drag acting on particles, the fluid threshold characterized by frictional wind velocity is derived from the balance of different equations for the forces acting on a particle [27, 28]. The constant frictional velocity can be conveniently obtained from the logarithmic profile of time-averaged wind speed in the lower portion of a fully developed turbulent boundary layer over a flat granular soil surface. Bagnold's framework has been regarded as the standard approach for predicting transport initiation, especially for medium-sized sand grains [29, 30]. Following Bagnold's seminal work, several refinements and improvements to his approach have been suggested, considering the effects of interparticle forces or cohesion forces, lift force and particle Reynolds number [2, 31-

33], and turbulence intensity [21, 34]. These corrections extend the application of Bagnold's model to a wide range of particle and fluid characteristics [35]. However, most of these advancements employ time-averaged flow measurements, failing to capture the highly fluctuating aerodynamic forcing leading to the intermittent and episodic character of natural movements of sediment particles generally taking place during turbulent airflows [18,36-40]. Analytical particle forcing models based on probabilistic approaches consider the effect of turbulent fluctuations [41, 42]; however, they fail to describe the rich dynamics of individual particles.

This study aims to diverge from the traditional focus of the literature away from the averaged wind grain forcing approaches, by adopting a novel particle dynamics approach to incipient grain entrainment due to instantaneously fluctuating aerodynamic forces. Such an approach should consider the different ranges of particle motion. Specifically, we use the term rocking, to define relatively small particle motions within its initial resting pocket, during which the particle may not get fully entrained (classified as a partial or incomplete entrainment, Figure 1). In the case of stronger (at or above the required threshold) instantaneous aerodynamic forcing, the particle will perform a downwind ascent over the ridge of the downwind particles forming its resting pocket, classified as a full or complete entrainment (see Figure 1).

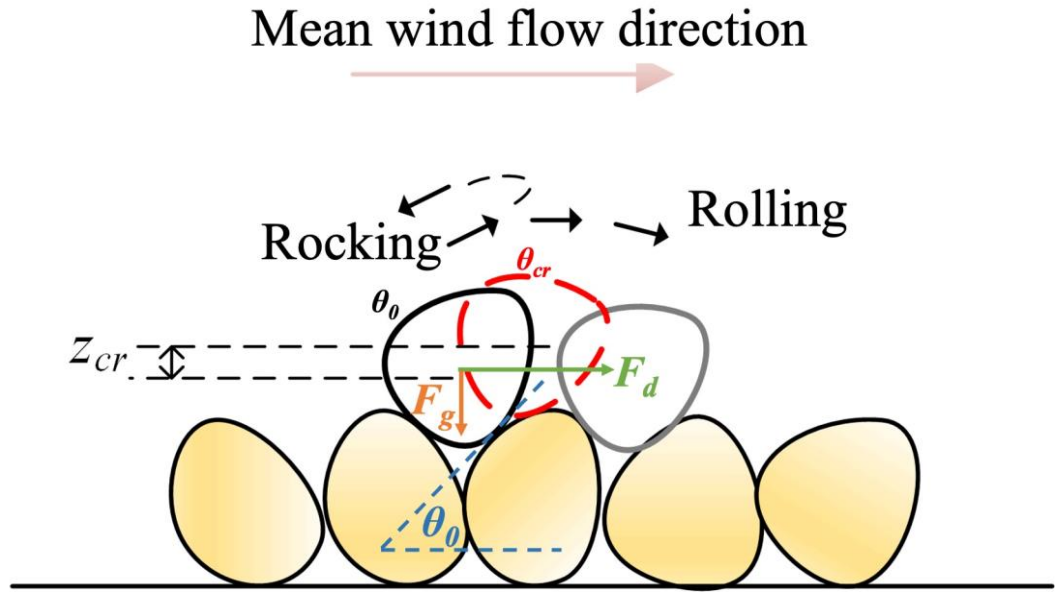

**Figure 1.** Sketch showing the classification of coarse particle dynamic motions over an aerodynamically rough surface, into rocking (incomplete entrainments where the particle falls back into its pocket due to short-lived aerodynamic forcing) and incipient rolling (full entrainments where the particle reaches the required critical angular displacement $\theta_{cr}$, corresponding to the minimum critical vertical displacement $z_{cr}$, from where it can be further entrained downwind). The entrainment is under the actions of the drag force ($F_d$) induced by the energetic airflow events and the resisting force $F_{cr} = F_g \tan\theta_0$, comprising the particle's weight ($F_g$) component.

These motions can generally be due to aerodynamic lift or drag or a combination of these. Herein the focus is on the role of aerodynamic drag, resulting on a grain's rocking or complete rolling motions. From experimental observations, it is noted that these grain motion events can be strong, referring to a higher rate of angular displacement, or weak, if the rate of displacement is lower. For the case of weak rolling, the resulting particle mobilization may just lead to a complete entrainment (where the particle only needs to reach the critical vertical displacement threshold $z_{cr}$, as explained in section 3). or may not suffice for this, in which case the particle falls back in its resting pocket (results in incomplete or partial displacements). The magnitude of instantaneous force alone may be insufficient to determine complete entrainment, even if it is above the predicted average threshold force for incipient motion [43]. This conceptual approach has been tested with bench-top and flume experiments [43, 44] and numerical simulations [45, 46]. Thus, this distinction between the end result of the effect of turbulent aerodynamic forcing, is necessitated if a truly dynamic criterion is to be accounted for. Further, the traditional spatially and temporally averaged approaches are useful under a traditionally implemented Eulerian framework. In contrast, an event-based approach might be better posed for a Lagrangian description of particle entrainments, which is particularly interesting for low transport conditions.

Here an initial wind tunnel survey is conducted to investigate the validity of the above framework in the context of aeolian transport. The instantaneous aerodynamic force is parameterized by the squared wind velocity ($u^2$) upwind of the test particle by assuming a constant drag coefficient ($C_d$) and properly selecting the representative frequency of wind velocity signal. The aerodynamic force may comprise low-frequency velocity fluctuations corresponding to larger turbulent eddies and the highly fluctuating turbulent energy component comparable to the particles' length scales. The average wind velocity ($\bar{u}$) in the initial survey is about 5 m/s, and particle diameter (*d*) is 40mm; then, the frequency matching the scale of eddies could be estimated by $\bar{u}/d$ equal to 125 Hz. [47] found that high-frequency flow components (> 10 Hz) have a negligible contribution. This is because high-frequency fluctuations in aerodynamic forces tend to be localized and vary rapidly across the particle's surface. For a large particle, these localized fluctuations often cancel each other out when integrated over the entire surface area. This spatial averaging effect means that the net force acting on the particle is primarily determined by lower-frequency, larger-scale flow structures rather than the high-frequency components. Also, coarse particles have relatively significant inertia, which means they respond much more slowly to force fluctuations compared to the timescales of high-frequency aerodynamic fluctuations. The particle's motion essentially acts as a low-pass filter, responding primarily to the time-averaged forces and lower-frequency fluctuations, while being relatively insensitive to rapid, high-frequency variations. This temporal averaging effect further diminishes the influence of high-frequency force components on the particle's overall motion. An example of time histories of $u^2$ (showing the low-pass signal both at 10 Hz and 125 Hz) and synchronous particle angular displacement ($\Delta\theta$) is shown in Figure 2. It is observed that both rocking and rolling of the test particle occur due to relatively less short-lived events of high magnitude of aerodynamic force ($\sim u^2$) above the initial threshold force ($\sim u_{cr,0}^2$; $u_{cr,0}$ is computed by the Bagnold's force model). Even though the magnitude of peak force during the rocking event is apparently greater than the magnitude of force during the rolling, at the same time, the latter is more sustained.

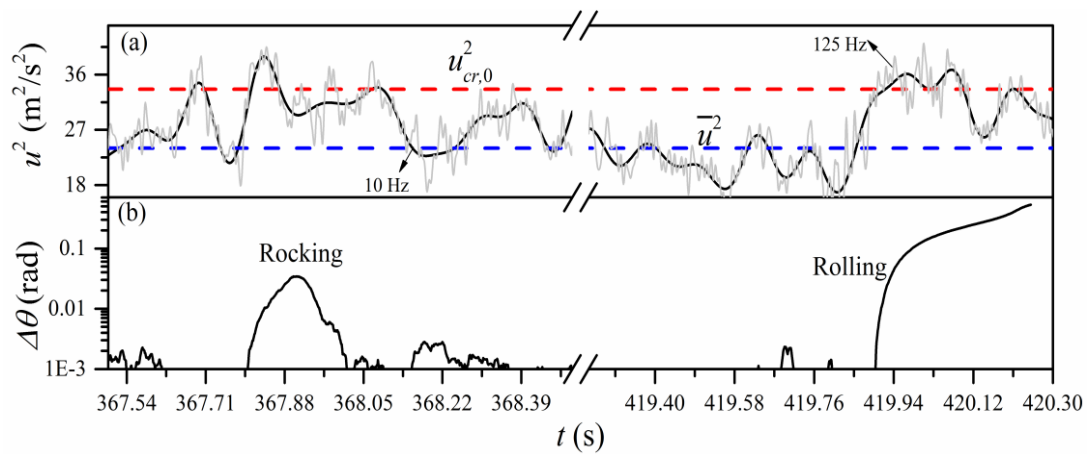

**Figure 2**. Example of synchronous time series of (a) aerodynamic force (parameterized by the square of windwise wind velocity upwind of the test particle, $u^2$. The grey and black lines refer to the quadratic representation of the low pass filtered velocity signal at 125Hz and 10Hz) and (b) angular displacement ($\Delta\theta$) of a test particle ($d$=40 mm and of 2.709 g weight) exposed fully on the solid surface at sub-threshold wind conditions, indicated by $\bar{u} < u_{cr,0}$.

Further, according to the theoretical time-averaged aerodynamic force-based models, entrainment could have occurred during any instances where the $u^2 > u^2_{cr,0}$. However, this condition is insufficient, as for relatively short-lived events, the particle's response may not be discernible (e.g., see $t$=367.7 in Figure 2). During such events, the local aerodynamic force exceeds the threshold; the particle may rock around a pivoting point, subsequently returning to the resting position eventually with no resulting out of pocket displacement (Figure 1). Conversely, if the event ($u^2 > u^2_{cr,0}$) lasts sufficiently long, the pivoting particle can escape from its resting pocket and has positive downwind displacement. Consequently, the traditionally employed time-averaged aerodynamic force criterion is not directly related to the critical vertical displacement ($z_{cr}$) separating rocking and rolling (Figure 1). In this manner, such criteria fail to differentiate between rocking and rolling and describe the change of particle displacement that is the focal point with regard to performing geomorphic work. It can be concluded that the magnitude of instantaneous aerodynamic force is an important criterion to differentiate between particles being at rest or in motion. However, alone it is not sufficient to describe the full entrainment.

## 3. Event-based energy model for entrainment of coarse particles by rolling

To address the challenges mentioned above, an energy criterion for aerodynamic entrainment of coarse particles is developed from an event-based energy model in this section.

## 3.1 Conceptual model

At near-threshold flow conditions, only a few intermittently occurring, short-lived and energetic wind bursts can overcome the intrinsic resistance of particles setting them into motion. An energetic flow event may impinge on a coarse particle resting on a aerodynamically roughened solid surface, as shown in Figure 3.

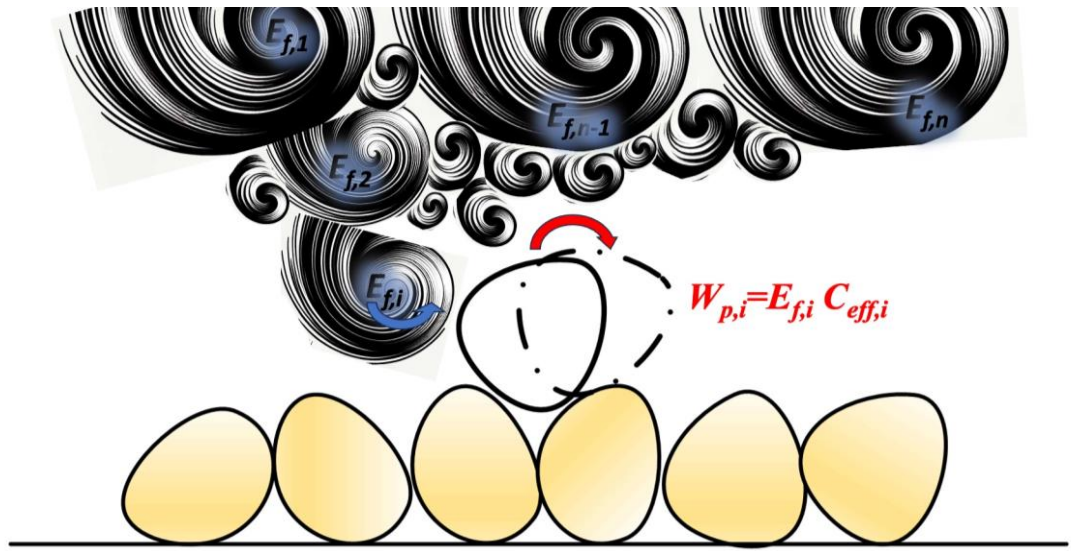

**Figure 3.** Sketch of the incipient entrainment of a downwind rolling particle due to a sufficiently energetic, turbulent flow structures. The mechanical work ($W_{p,i}$) done on the particle due to the impinging flow structure (with energy content $E_{f,i}$), depends on the efficiency of energy transfer $C_{eff,i}$. The turbulent air flow is now represented with an equivalent spatially evolving distribution turbulent airflow structures of varying length scales and energy intensities, $E_{f,i}$, for i =1 to n.

Neglecting interparticle forces and considering energy conservation, the energy content of the flow event, $E_f$ is converted into the mechanical energy of the particle with a transfer coefficient $C_{eff}$ (similar to [24]). To overcome the energy barrier resulting from the critical vertical displacement $z_{cr}$, as shown in Figure 1, the energy offered from the flow event ($C_{eff}E_f$) should be at least greater than the increase of geopotential energy during the dislodgement ($F_g z_{cr}$; $F_g$ is the effective gravity of the particle). The above analyses yield:

$$E_f \geq F_g z_{cr} / C_{eff} \tag{1}$$

where the equal sign denotes the threshold event energy ($E_{f,cr}$). Eq. (1) presents an energy criterion where the character of flow events and full entrainment of particles are interlinked at the energy level.

### 3.2 Theoretical definition of energetic flow events

Sufficiently energetic aerodynamic events that transfer energy to particles are referred to as energetic airflow events. Energy transfer is achieved by means of the airflow structures performing aerodynamic work on the entrained particle.

It is hypothesized that a flow event passes over a resting coarse particle exposed on the solid surface results in an episodic rolling, as shown in Figure 3. The flow event refers to a translating flow structure assumed to be frozen according to Taylor's frozen turbulence hypothesis. The windwise forces considered here include drag force ($F_d$) induced by the flow event and surface resistance ($F_{cr}$), which consists of the particle's weight component along the direction of motion. The effects of cohesion are negligible for coarse particles, but the proposed framework can still be applied even for small particles, by considering an added parameter to the resistance force, $F_g + \beta\,d$, where $\beta$ is a dimensional parameter of the order of $10^{-5}$ N/m, as shown in [31]. In the following analysis of experimental results, the above cohesion term is negligible thus it is not further considered. The particle's windwise movement begins at the time $t_0$ and ceases after one or several acceleration-deceleration cycles because of the variation of $F_d$ and $F_{cr}$ along the rolling path. Conversely, airflow components cannot perform mechanical work when the resisting forces are greater than the driving forces, resulting in the particle remaining at rest and no energy transfer happening towards its entrainment. Therefore, the portion of the flow event responsible for incipient rolling, as shown in Figure 3, will be dependent on the resistance to motion features and can be described as follows:

$$F_d(t_0) > F_{cr} \text{ and } \int_{t_0}^{t_0+T_f}(F_d - F_{cr})dt = 0 \qquad (2)$$

The first term of the above equation indicates such flow events are initiated at $t_0$ when the temporal aerodynamic force $F_d(t_0)$ exceeds the initial resistance level $F_{cr,0}$ (resistance level at $t_0$) sets the particle into motion. The second term determines the duration of the aerodynamic forcing event $T_f$, i.e., the elapsed time from the instant the particle begins to move until the instant where the particle has lost its offered momentum. This generic definition allows one to analyze all cases where energy is offered towards performing mechanical work regardless of the magnitude of the particle displacement. This way all types of particle motion can be considered, from creeping and small particle displacements (rocking motions), to rolling motions towards complete entrainment and beyond that (for incipient saltation for even stronger energetic events or different types of energetic events than assessed here e.g., due to impulsive aerodynamic lift). Here, the duration of an aerodynamic flow event $T_f$, considers the sum of all consecutive periods (after the instance where the temporal aerodynamic force $F_d(t_0)$ exceeds the initial resistance level $F_{cr,0}$), where the instantaneous aerodynamic forcing is greater and lower than the resisting forces, thus offering momentum towards the particle's full entrainment and expending the particle's momentum, respectively, until the particle's momentum is exhausted. This is a broader, result-oriented definition and different to definitions of energetic flow events given in the past [24, 44]. At the limit case of incipient motion, the sum of momentum given by the impulsive forcing and taken from the particle due to the resisting forces shall be equal to zero,

The above analysis, without loss of generalization, does not take into account incipient entrainment by saltation, since it is difficult to parametrize lift forces acting on particles very near the soil surface [48, 49]. Even though this may be important to consider at times [48], there exist difficulties in accounting for lift forces due to their strong and poorly understood decrease with vertical distance from the surface pocket [49]. One can equally consider the effect of a dominant (drag or lift) or total aerodynamic force, but here we only consider drag, aiming for a simplified yet well working definition (as later demonstrated by the final agreement between the observations and predictions, which can be seen as a validation of the involved approximations) of a criterion for the duration of aerodynamic forcing event (one that can also be applied to experimental data, as shown in section 5).

Eq. (2) is also suitable for characterizing energetic flow events responsible for rocking because the downwind ascent of rocking particles beginning at the particle's resting position and ceasing at the rocking peak follows a similar dynamical process, as the rolling shown in Figure 1. Using the traditional quadratic drag force parameterization, $F_d \sim u^2$, $F_{cr} \sim u^2_{cr}$, we obtain the conceptual definition of energetic flow events described by wind velocities:

$$u^2(t_0) > u^2{}_{cr} \text{ and } \int_{t_0}^{t_0+T_f}(u^2 - u^2{}_{cr})dt = 0 \qquad (3)$$

Eq. (3) could be regarded as an axiomatic definition for defining the duration energetic flow events based on Newtonian mechanics. However, its direct applicability is hindered analytically because the resisting force depends on the particle's location, thus, varies as the particle gets entrained. An alternative approach is to consider the resistance force approximately constant and equal to the resistance estimated considering the initial particle configuration, resulting in replacing $u^2_{cr}$ with $u^2_{cr,0}$, in deriving the following more practical definition:

$$u^2(t_0) > u^2_{cr,0} \text{ and } \int_{t_0}^{t_0+T_f}[u^2 - u^2_{cr,0}]dt = 0 \qquad (4)$$

The operation of Eq. (4) is demonstrated in Figure 4a, where flow events responsible for both rocking (e.g., event A) and rolling (e.g., event B) are extracted in terms of wind velocities measured upwind of target particles. The starting and end time for extracting the flow events are defined by Eq. (3). However, the critical resistance levels estimated while considering the initial particle position shown in Eq. (4) may result in the extracted flow event durations being shorter than measured. This is so because for slowly dislodging particles, the resistance reduces as the particle gets more exposed, rendering more of the forcing flow event effective in transferring energy to move the particle. For example, by the end of event B, full particle entrainment (denoted with the black dashed line) has not been achieved, because the energy transferred from the forcing flow event (see grey shaded area for event B above $u^2_{cr,0}$, in Fig.4), that needs to be balanced from the work of resisting forces as the particle completes its full displacement (see grey shaded area for event B below $u^2_{cr,0}$, in Fig.4), is greater than what is estimated considering fixed resisting forces, while also the latter is also offered at a decreasing rate. Further discussions of the influence of surface resistance level on the identification of energetic flow events will be presented in Section 5.1.

With the aid of Eq. (4), the feasibility of the energy criterion provided by Eq. (1) can be demonstrated in terms of energy densities ($\sim u^3$) by comparison of the shaded areas, $E_{f,A}$ and $E_{f,B}$ (Figure 4b) if the energy transfer coefficient is constant. Rolling should be larger than the threshold energy ($E_{f,cr}$) and the opposite for rocking.

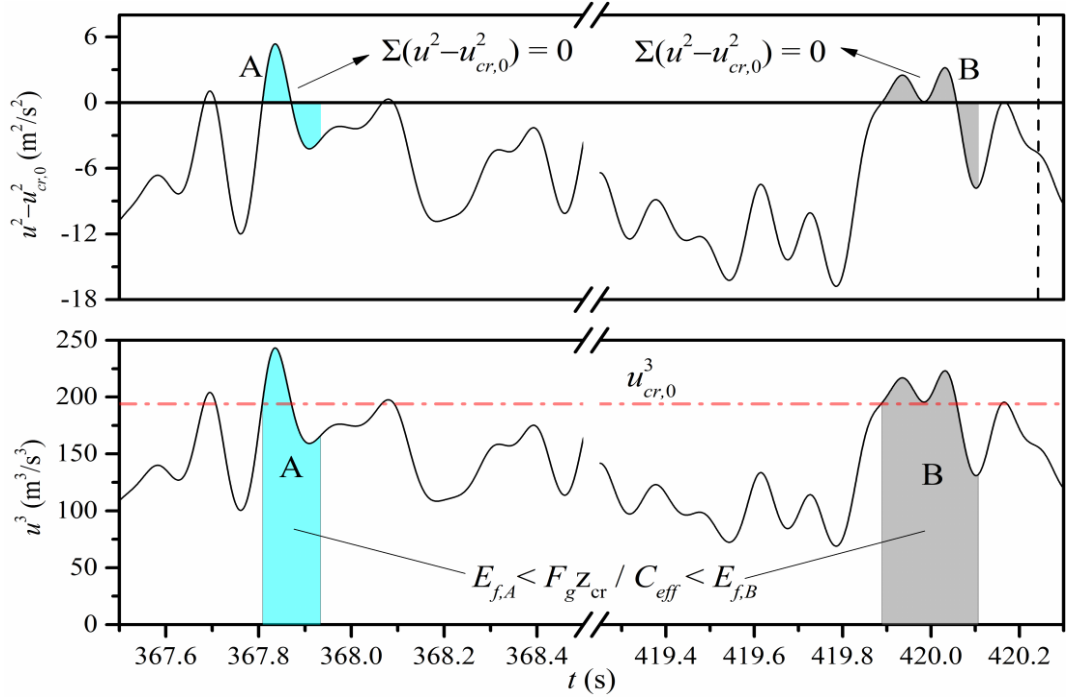

**Figure 4.** Top: Extraction of the energetic flow events according to Eq. (4). Bottom: Distinct event energy for the case of particle's rocking and the rolling (A and B, respectively, shown in Figure 2) in terms of wind velocity signals upwind of particles. The black dashed line denotes the moment at which full entrainment is achieved.

### 3.3 Energy criterion for coarse particles exposed on a uniform surface of identical spheres

In the heterogeneous soil surface shown in Figure 1, $z_{cr}$ will not be a constant but a stochastic value. However, assuming spherical particles spread loosely over a uniform soil surface, there is no loss of generality in using the ideal grain arrangement shown in Figure 5 to develop $z_{cr}$, which is a function of initial angular position ($\theta_0$), critical angular position ($\theta_{cr}$) and length of the rotation axis ($L_R$):

$$z_{cr} = L_R(\sin\theta_{cr} - \sin\theta_0) \tag{5}$$

Further, event energy ($E_f$) is defined as the following integral [50], which also introduces the drag coefficient ($C_d$):

$$E_f = 0.5\rho_a C_d S \int_{t_0}^{t_0+T_f} u^3 \, dt \tag{6}$$

where $\rho_a$ is the air density, and $S$ ($=0.25\pi d^2$) is the cross-sectional area of spherical particles. Combining Eqs. (1), (5), (6), and considering the particle's weight force $F_g=(\rho_p-\rho_a)\pi d^3 g/6$, yields the following energy criterion, for spherical particles exposed on uniform soil surfaces:

$$\int_{t_0}^{t_0+T_f} u^3 \, dt > \omega^2 L_R (\sin\theta_{cr} - \sin\theta_0)/C_{eff} \tag{7}$$

where $\omega (= \sqrt{4(\rho_p - \rho_a)gd/(3C_d\rho_a)})$ is the terminal velocity of particles in air [51]; $\rho_p$ is the particle density, and $g$ is the gravitational acceleration. The settling velocity appears in Eq. (7) as it parametrizes the windwise drag force. The latter can be expressed as the ratio of the work of drag force to the energy content of flow events [24]:

$$C_{eff} = \left[\int_{t_0}^{t_0+T_f} u^2 v \, dt\right] / \int_{t_0}^{t_0+T_f} u^3 \, dt \approx \bar{v}/\bar{u} \tag{8}$$

where $\bar{u}$ and $\bar{v}$ are the windwise flow and particle velocity, respectively, averaged during the duration of the flow forcing events (effectively over the duration where the aerodynamic force keeps supplying energy for the displacement of the grain).

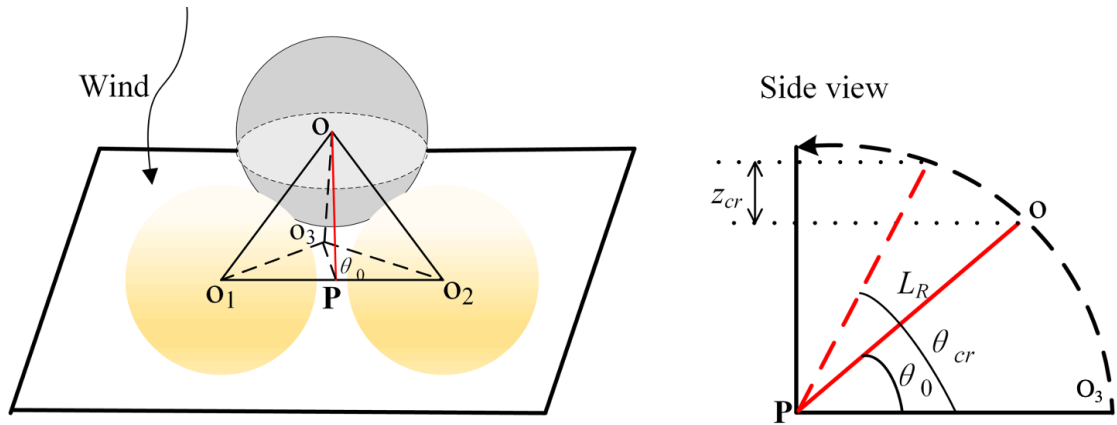

**Figure 5.** Sketch of the three-dimensional pocket geometry, looking at the particle configuration from the downwind direction, to illustrate the critical elevation ($z_{cr}$) for entrainment by rolling mode for a spherical particle over a well packed uniformly roughened surface. The wind direction is towards the reader, pushing the particle over the downwind path over the saddle formed by the two downwind grains forming its resting pocket.

## 4. Experiments

To examine the energy model for incipient aerodynamic entrainment of coarse particles presented above, a series of well-controlled experiments were performed in an aspirated environmental wind tunnel in the State Key Joint Laboratory of Environmental Simulation and Pollution Control at Peking University. The available section of this tunnel is rectangular and 30 m long, 3 m wide and 2 m high. The temporal average of the wind velocity measured in the middle of the wind tunnel cross-section, 0.4 m away from the tunnel entrance and 1.2 m above the tunnel floor, was used as the characteristic reference velocity ($U$) to denote the magnitude of airflow used in the experiments. $U$ can be adjusted continuously in a range of 0.5 to 20 m/s. Note that the choice of such a relatively large wind tunnel for our experiments was motivated by the corresponding relatively large boundary layer, which allows for larger peak fluctuations of the wind velocity relative to its mean value and, thus, a larger frequency of energetic flow events [25, 52-54].

In real-world cases of wind-induced incipient entrainment of coarse sediment particles, the variability of coarse grains and their local arrangement will affect aerodynamic and resistive forces. The complexities associated with grain and surface variability were removed to reduce the phenomenon to its essential features by using hollow spherical particles of *d*=40 mm and weighting 2.709 g (e.g., typical table tennis ball, made of celluloid), both for the test particle and the uniform and non-erodible surface arranged in the square array (Figure 6). If natural gravel of a similar size to the test particles were used in these experiments, a high-speed tunnel capable of airflows up to 100 m/s would be required [13]. Thus, the low density of the particle (about 67 times the density of air) is chosen to allow running the experiments in the available low-speed wind tunnel [55]. The modelled surface is 4 m long and 2.4 m wide, located 22.6 m downwind from the tunnel entrance.

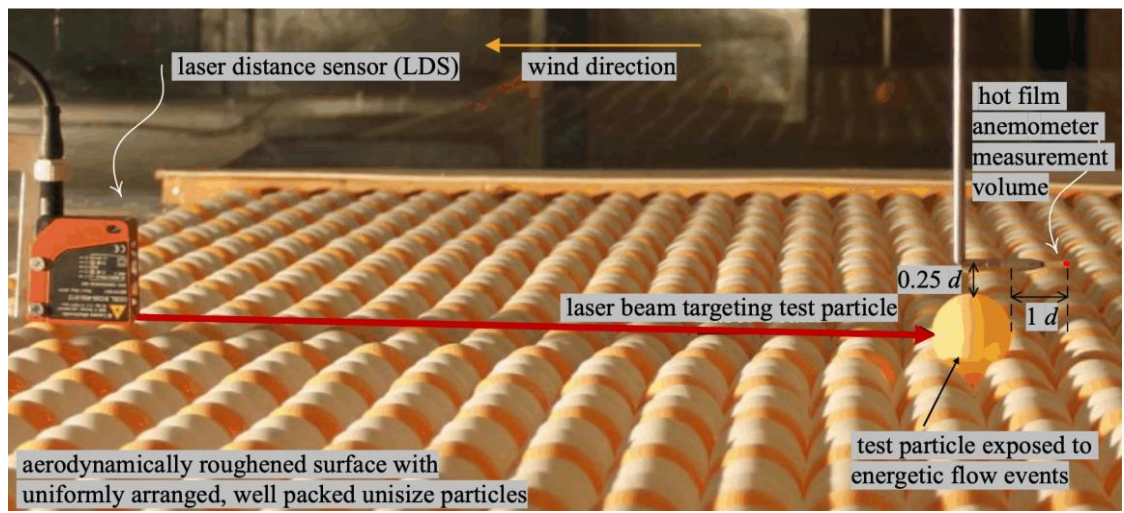

**Figure 6.** Illustration of the experimental setup demonstrating the laser distance sensor for recording the horizontal displacement of the exposed test particle and a hot film anemometer wind velocities upwind of the particle using. The airflow direction is from right to left.

A 0.1 mm-resolution laser distance sensor (LDS, manufactured by Leuze Electronics, Germany) was used to track the motion of the target particle. A red laser beam with a 655 nm wavelength emitted from the LDS shone on the center of the target particle. Once the particle began to rock along or against the wind direction, the sensor could measure the corresponding displacement. The sensor's measuring range was set from the rest position of the target particle to the local high point where the mobile particle would be considered to have been completely entrained. The setup is calibrated as demonstrated in Xiao-Hu et al. [55]. Moreover, the measured displacement can be transformed into the corresponding angular displacement based on the pocket geometry.

The instantaneous wind velocities upwind of the target particle ($u$) were recorded using a V-hot film anemometer manufactured by DISA Elektronik in Denmark. The anemometer was fixed at $1d$ upwind of the front of the particle and $0.25d$ above the top of the particle (Figure 6). The velocity measurement location was chosen to avoid being too close to the particle, potentially interfering with the local flow field around the particle, or too far where the measurements would not represent the local aerodynamic forcing field around the test particle.

The particle displacements and wind velocities sequence signals were collected simultaneously using a multichannel signal processor with a data acquisition rate of 1000 Hz. First, the raw signal of particle displacement was smoothed using a 50-point-adjacent-averaging method to remove high-frequency noise, such as the tunnel rocking and electromagnetic interference from other instruments. Next, the measured time series of wind velocities were filtered using an FFT low-pass filter with a 10 Hz cutoff frequency (following the rationale presented in section 2). The filter first transformed the time series of wind velocities to the frequency domain employing the Fast Fourier Transform (or FFT) and then set the amplitudes for frequencies > 10Hz to zero; finally, it converted the result back to the time domain with the inverse FFT.

The mean airflow velocity profile at the measuring location follows the typical logarithmic law 10 mm above the tops of the particles comprising the modelled surface (Figure 7a). The turbulence intensity decreases with the height from the modelled surface, with a value of up to 16% near the surface, indicating the presence of increased shear forces on average (Figure 7b). The aerodynamic roughness estimated from the logarithmic wind velocity profile is approximately 1/8 of the particle diameter.

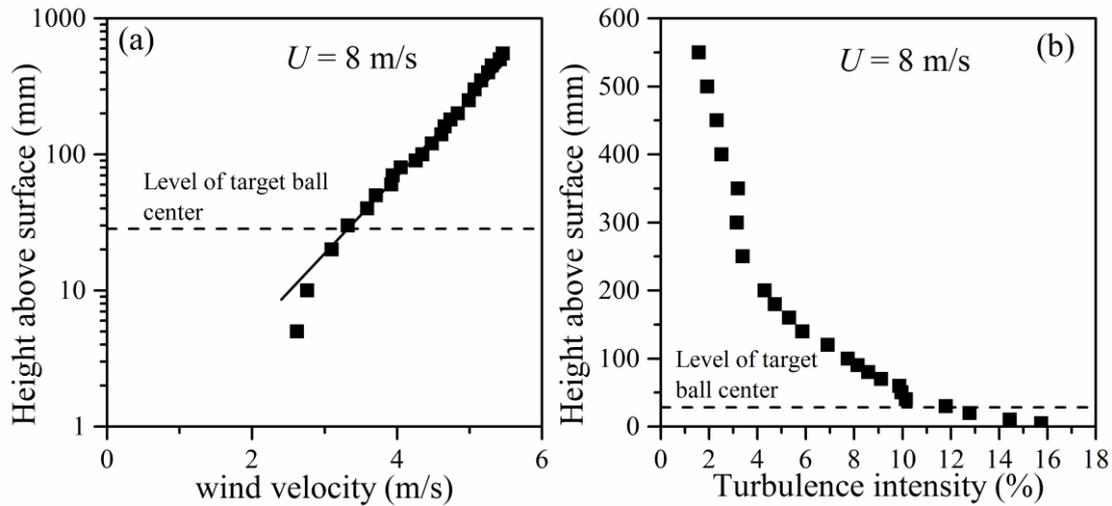

**Figure 7.** Indicative profiles of wind velocity and turbulence intensity at the measuring location for U = 8 m/s. The zero-height reference level is set as the tops of the spherical particles comprising the rough surface. (a) Time averaged wind velocity profile. (b) Profile of the measured turbulence intensity.

The particle could hardly move at $U < 7.5$ m/s, while at $U > 8.2$ m/s, the target particle would rapidly roll away from the measurement position soon after it was set in its resting pocket. Thus, the experimental wind strength, representative of near incipient flow conditions, increased from 7.5 m/s to 8.2 m/s. The sampling durations were limited to the total time required for complete entrainment of the test particle further downwind and out of the range of the laser distance sensor (measurements were interrupted after 900 seconds if full entrainment had not occurred, and this only happened once). For the range of airflows tested above, sampling durations varied from a few seconds to several minutes. Following the presented methodology, twenty-four runs (E1-E24) were conducted representative of different (increasing velocity) but near incipient motion flow conditions (Table 1). During the first four runs (7.5-7.8 m/s), the particle only performed minute displacements and rocking events. During the rest of the runs (7.9-8.2 m/s), incomplete (rocking motions) and complete entrainment (involving rolling out of the local pocket) of the test particle were observed. Even though the reference mean velocity ($U$) and frictional velocity ($u_*$) are about the same for all these experiments, the average frequency or rate of occurrence of rocking ($f_{rc}$) and rolling ($f_{rl}$), is observed to change more than an order of magnitude (Table 1).

**Table 1** Summary of flow characteristics for incipient entrainment experiments

|  | $U$ (m/s) | $u_*$ (m/s) | $Re_*$[a] | $f_{rc}$ (count/s) | $f_{rl}$ (count/s) |
|---|---|---|---|---|---|
| E1-E4 | 7.5-7.8 | 0.259-0.269 | 691-717 | 0.00556-0.0344 | 0 |
| E5-E24 | 7.9-8.2 | 0.281-0.304 | 749-811 | 0.0689-0.299 | 0.00143-0.194 |

[a] $Re_*$ ($= du_*/\nu$) is the particle friction Reynolds number, and $\nu$ ($= 1.5 \times 10^{-5}$ m$^2$/s) is the kinematic viscosity of air.

## 5. Results and discussion

### 5.1. Comparison of measurements to the theoretical definition of energetic airflow event

As mentioned in section 3.2, the theoretical definition of energetic flow events is associated with $u^2_{cr}$. For practical purposes, the time-independent initial resistance force is considered representative of the resistance to motion throughout the application of aerodynamic *forcing ($u^2_{cr}$* in the second term of Eq. (4) is assumed to be ~ $u^2_{cr,0}$). This critical level can be computed using Bagnold's force model, $u^2_{cr,0} = F_g\cos\theta_0 / (0.5\rho_a S C_d \sin\theta_0)$. The windwise quadratically parameterized threshold, $u^2_{cr}$, is concerned with the time-dependent displacement of particles [56]. An example of conceptually demonstrating this is presented in Figure 8. The movement in the figure is the strongest rocking (critical rolling) among the 568 rocking events we detected. It is initiated by the instantaneous drag force (~ $u^2$) exceeding the initial resistance force. As the particle rolls towards the saddle formed by the downwind particles, $u^2_{cr}$ decreases until the particle reaches the peak where the minimum resistance level ($u^2_{cr,m}$) occurs (Figure 8a). The replacement of the time-dependent $u^2_{cr}$ with $u^2_{cr,0}$ overestimates the average resistance force (~$u^2_{cr}$) or total resistance force during the downwind ascent. The overestimation results in that the event (shadow in Figure 8a) predicted by Eq. (5) is shorter than the measurement, corresponding to the flow segment during the downwind ascent. Specifically, the prediction is a fraction of the measurement. Such aerodynamic forcing duration shortening effects are obvious for rocking motions that are slowly occurring, such as the rocking event shown in Figure 8, but not significant for impulsively forced rocking motions that take place fast (e.g., event A in Figure 4). This influences the efficiency of energy transferred from the flow events to particles ($C_{eff}E_f$). Such influence for the strong rocking is demonstrated in Figure 8b, which shows that the predicted energetic flow event does not account for all but only a part of offered energy responsible for the downwind ascent.

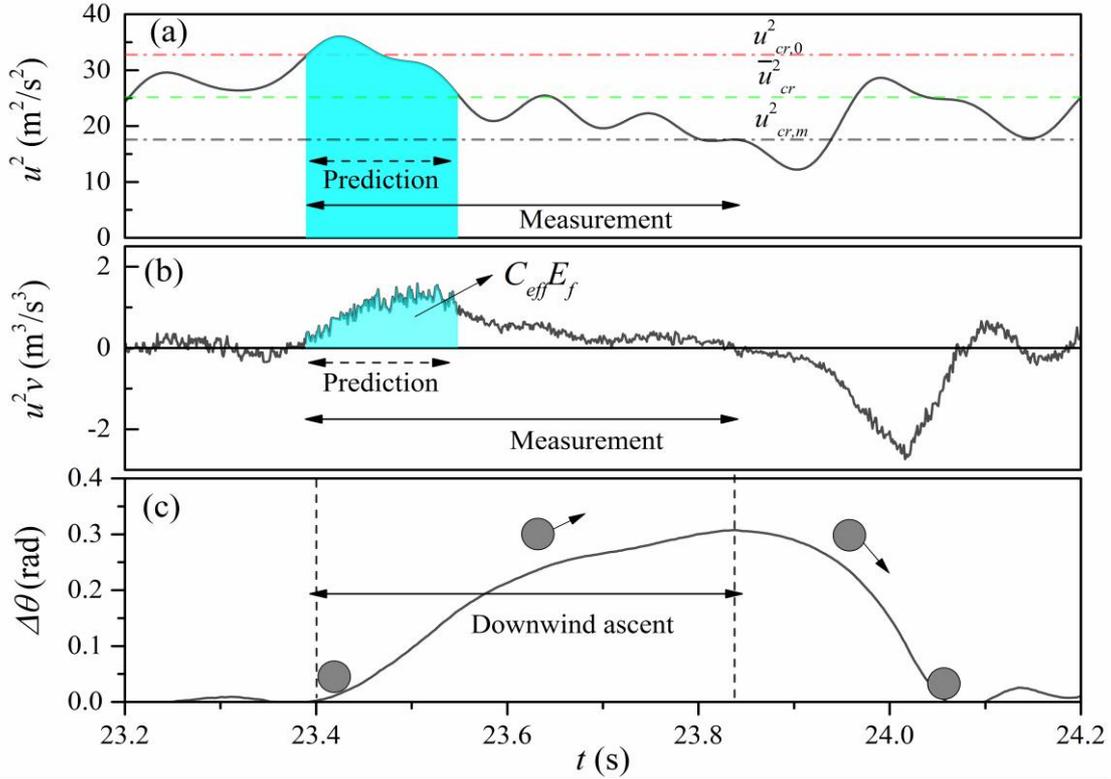

**Figure 8.** Synchronous time series of (a) drag force (~ $u^2$), (b) rate of energy transfer (~ $u^2v$) and (c) particle angular displacement ($\Delta\theta$) near the instant of the strongest rocking (critical rolling). The figure is intended to compare theoretical energetic airflow events (shadow) predicted by Eq. (4) with measurements.

The analytical predictions and experimental measurements of the energy offered to the particle are positively correlated, as shown in Figure 9. The scale factor approaches 1 for the weak rocking of low amplitude (less than $0.04\pi$ in our experimental conditions) and is between 0.62 and 1 for relatively strong rocking motions. It decreases with the increase of rocking amplitude, which results in a higher average exposure level of particles and subsequent greater difference between average and initial resistance levels leading to shorter predicted events than measurements. Because of the positive relationship between offered energy, we could use the predicted events to represent the measured events for discerning particle dislodgement.

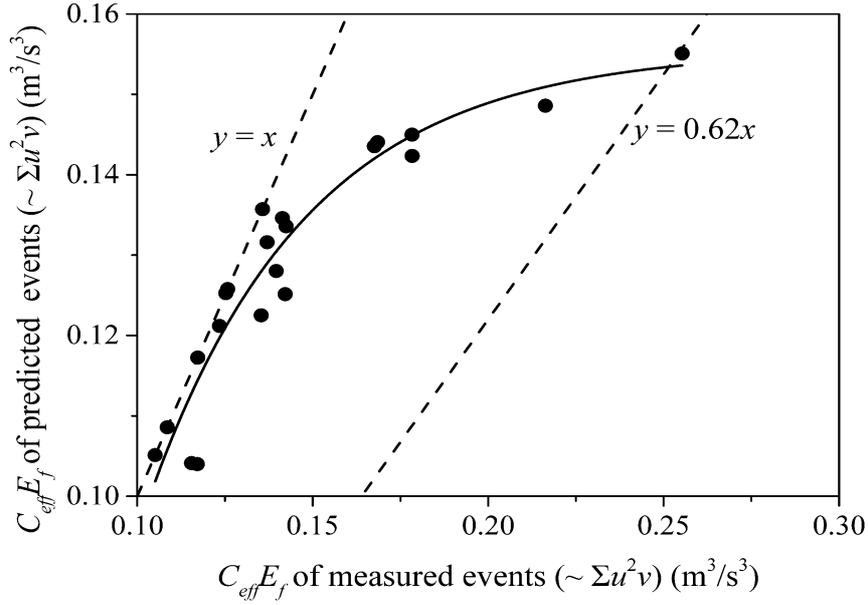

**Figure 9.** Comparison of the analytically predicted ($C_{eff}E_f \sim u^2v$) and experimentally estimated energy offered to the coarse particle from the energetic flow events, resulting in relatively strong rocking (with angular amplitude greater than $0.04\pi$).

The event-based criterion is shown herein to have the potential to allow the useful distinction between the flow components that are or are not capable of entraining coarse particles, which is foundational for modelling intermittent grain transport in near-threshold turbulent flows [57, 58]. Some attempts have been made to model sand grains' aeolian transport due to coherent flow structures, typically bursting-like events [59-61]. However, such progress is relatively absent for coarser particles. This study develops a novel approach to identifying flow events accounting for the episodic movements of coarse particles, containing rich particle dynamics that are insufficient in detection of bursting-like events (e.g., see, for example, the quadrant technique of Lu and Willmarth amongst others [62-64]). As such, it may be considered a new paradigm for aeolian transport.

## 5.2. Comparison of measurements to theoretical predictions of energy criterion

The energy criterion defined by Eq. (7) is available once undetermined parameters, including $\theta_{cr}$ and $C_{eff}$ are quantified from experiments. Based on the obtained particle angular displacement, the maximum angular location of the particle pushed by energetic flow events ($\theta_{cr}$) is about $0.36\pi$ except in incipient rolling. The value is obtained by the sum of the initial angular position ($\theta_0=0.3\pi$) and angular displacement flow events achieved in the critical rolling (see angular displacement at the right boundary of shadow shown in Figure 8c). This might be true because once the particle moves beyond a specific position, the mean flow forcing greater than the reduced resistance is enough to entrain the particle completely [65].

Estimated $C_{eff}$ results are shown in Figure 10. It is generally observed that distinct $C_{eff}$ values are found to correspond to rocking and incipient rolling motions. Starting from the lowest $C_{eff}$ values for which motion was observed, these correspond to levels of about $C_{eff} \sim 0.00048$ (Figure 10). This can be thought off as the threshold for the occurrence of creep [15], which is both relatively infrequent as well as demonstrably of very low efficiency. At a twice this level of $C_{eff} \sim 0.00096$, the particles perform clear rocking motions (yet incomplete to entrain the particle downwind). These events, all the way up to efficiency levels of about an order of magnitude greater than the creep threshold, can be thought to correspond to rocking motions that could contribute to mechanical sieving processes, that are essential for the formation of megaripples under certain conditions. This threshold $C_{eff} \sim 0.0048$, can be considered as the required threshold for incipient rolling of coarse particles (Figure 10). For larger forcing events, rolling will be more energetic, and for the cases assessed herein it was seen that the rolling could not be sustained for higher efficiency levels. For example, one could consider the level of $C_{eff} \sim 0.0096$ to correspond to the critical saltation threshold. This level is twice as much as the critical rolling threshold and an order of magnitude greater than the threshold for rocking (Figure 10). These transitionary levels are obviously not crisp thresholds but can rather be seen as relatively fuzzy boundaries, which can be used as a means to classify the distinct resulting motions and their contribution to the respective continuous transport processes. Specifically, for the highest magnitude and relatively short-lived aerodynamic forcing events, even though the particle is seen to start being displaced by rolling it then fast makes a hopping motion and gets further transported downwind with stronger such saltation motions. As the energy content of these turbulent flow structures would further increase, the initial rolling motion would become even shorter and eventually non-discernible, initiating entrainment in saltation mode. The highest efficiency entrainment events e.g., those with $C_{eff}$ greater the above defined saltation threshold (~0.0096), would correspond to almost saltating grains.

These thresholds can be drawn as a guide to the eye in Figure 10. One can distinguish the thresholds of creep, rocking, rolling and saltation, for the $C_{eff}$ values of 0.00048, 0.00096, 0.0048, and 0.0096, respectively. These thresholds can in turn be used to refer to the regimes of small vibrational motions (leading to creep), rocking motions (leading to mechanical sieving processes), and rolling or saltation motions (referring to incipient entrainment by the corresponding modes).

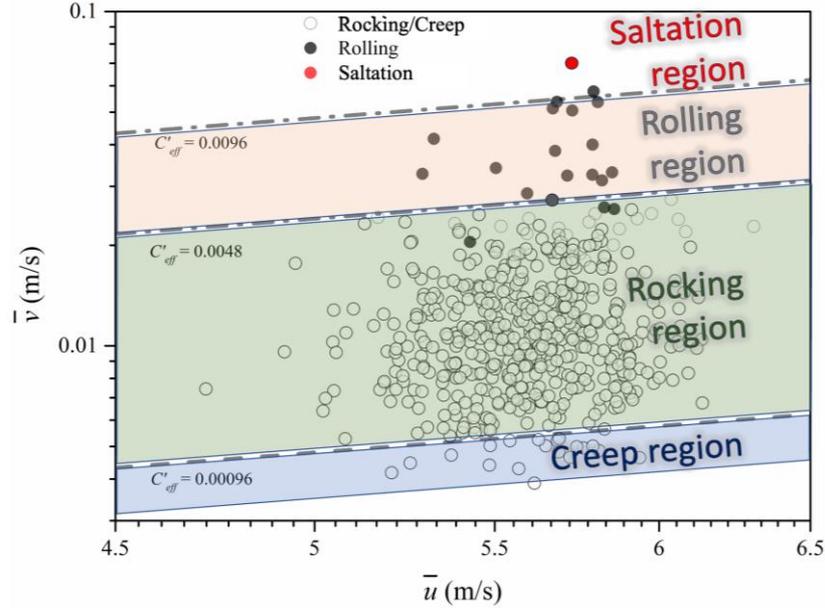

**Figure 10.** Plot of average wind velocities ($\bar{u}$) upwind of the test particle against average particle windwise velocity ($\bar{v}$) during energetic flow events (Eq. 4) resulting in particle movements. The data are from all experiments. The approximate threshold energy transfer efficiency lines are drawn as a guide to the eye, to allow distinguishing the different regions of creep, rocking, rolling and saltation, for increasing $C_{eff}$.

The events defined by Eq. (4) resulting in particle movements are extracted for analysis. For the case of small rocking motions, where the particle stays within its pocket, all events have displacements that are well captured by and fall within the measuring range of the LDS probe. However, for the case of rolling and complete entrainment motions, some of the events may result in particle displacements that cannot be captured by the LDS. Likewise, as the particle gets to fully entrain, it is apparent that the growing distance from the fixed hot wire anemometry probe will render the aerodynamic forcing assessment less representative to the forcing actually acting on the downwind moving particle. In case the flow forcing event has durations comparable to the full entrainment duration (e.g., see Figure 11b), the confidence in these measurements decreases. For practical considerations, and to reduce the influence of increased uncertainty data in the assessments in these cases, the duration of the flow structure is considered to match the duration of particle full entrainment out of its pocket.

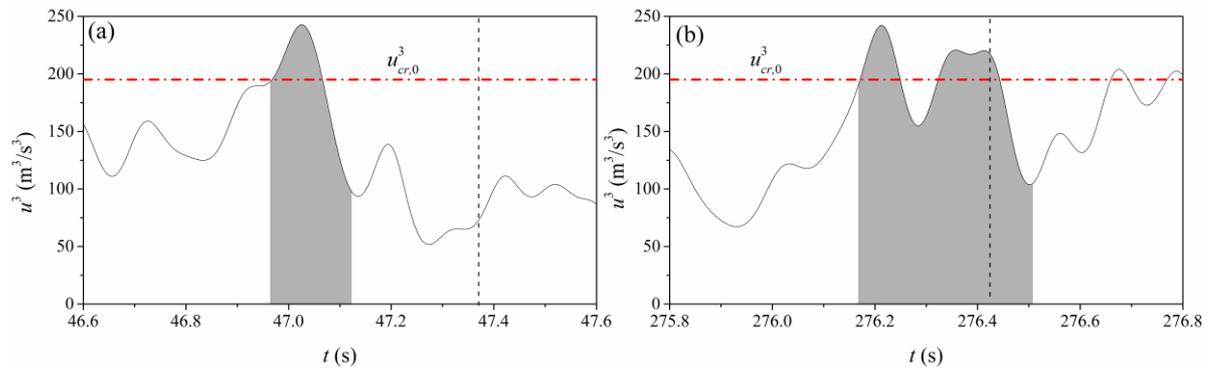

**Figure 11.** Short and long energetic flow events defined by Eq. (4), for rolling: (a) for impulsive or aerodynamic forcing events that are short-lived relatively to the particle's entrainment time, the assessed flow event is expected to well represent the forces acting on the particle; (b) for relatively long-lasting aerodynamic forcing events of durations comparable to the duration of particle's full entrainment, the downwind spatial separation of the particle from the fixed local anemometry measurement volume, may result in an increased relative error for the assessment of the actual energetic flow structure, responsible for the continued particle's response. It is noted that the long event comprises more than one cycle of strong-and weak phases. Black dashed lines represent the moment when full entrainment is achieved.

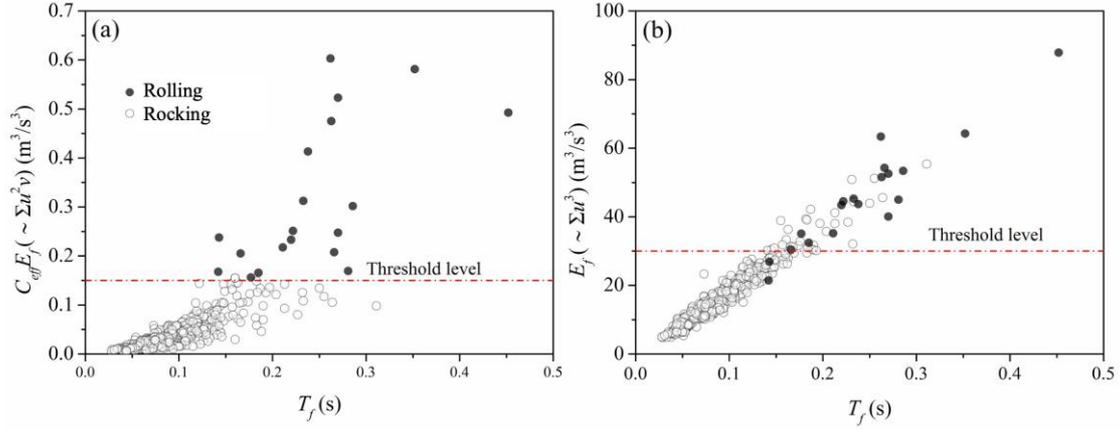

**Figure 12.** Comparison of predictions of the energy criterion defined by Eq. (7) in terms of (a) offered energy ($C_{eff}E_f \sim \Sigma u^2 v$) and (b) energy content ($E_f \sim u^3$) of energetic flow events defined by Eq. (4). Parameter values related with properties of the flow, the particles, and their configuration: $\rho_p$ = 80.8 kg/m$^3$, $\rho_a$ = 1.205 kg/m$^3$, $g$ = 9.8m/s$^2$, $C_d$ = 0.76 [66], $\theta_0 = 0.3\pi$, $d$ = 40 mm and $L_R = d\sqrt{3}/2$. The data are from all experiments.

Theoretical predictions of the energy criterion defined by Eq. (7) to lab measurements are shown in Figure 12, respectively, in terms of offered energy ($C_{eff}E_f$) and energy content ($E_f$) of flow events. The reliability of the presented criterion is supported strongly by the result shown in Figure 12a, indicating that full entrainment is achieved by energetic flow events capable of imparting sufficient energy to particles. The predictions shown in Figure 12b according to threshold energy quantified using a constant $C_{eff}$ (=0.0048) become less accurate; however, it accounts for approximately 89% of incipient rolling and 95% of rocking motion (Figure 12b).

### 5.3. Application of energy criterion for the aeolian transport of coarse particles

The rocking frequency $f_{rc}$ of coarse sediments may be one of the relevant parameters in modelling aeolian megaripples. Our studies indicate particle rocking arises from the action of energetic flow events, which are required to possess less energy than the threshold level. Then $f_{rc}$ could be estimated from the frequency of flow events ($f_f$):

$$f_{rc} = f_f P(E_f < E_{f,cr}) \tag{9}$$

where the symbol "$P$" denotes probability; $P(E_f < E_{f,cr})$ is the occurrence probability of rocking.

Further, the creep of coarse particles is vital for the evolution of coarse-grained bedforms. For example, the migration of coarse sediments on the crest of aeolian megaripples may limit the growth of megaripples [67, 68]. Therefore, the rate of aerodynamic dislodgement may be required as the initial input for modelling these bedforms. Following the energy criterion, it may be estimated by:

$$N_p = N_f P(E_f < E_{f,cr}) \tag{10}$$

where $N_p$ is the count of fully-dislodged particles per unit width and unit time; $N_f$ is the count of flow events per unit width and unit time; $P(E_f > E_{f,cr})$ denotes the occurrence probability of rolling.

## 5.4. Next steps towards monitoring and assessing the potential for geomorphic work due to turbulent wind flows

The findings from this study represent a significant advancement in assessing the probability of aerodynamic entrainment due to turbulence. These criteria and methods can also be used beyond the lab, indirectly by deploying high resolution anemometry or directly by using innovative assessment tools, as already has been demonstrated for fluvial environments (see [69-71]). This research builds upon recent conceptual and theoretical progress in the field of coarse material transport in turbulent flows, both in aquatic and aerial environments. Currently there exists the ambition to bridge the gap between laboratory insights and real-world applications by leveraging cutting-edge sensor technologies. These new sensors, capable of measuring minute particle displacements with unprecedented precision, allow deducing entrainment probabilities based on the frequency and magnitude of these micro-movements (similar to what has been identified and analyzed in the above sections).

Such technological breakthroughs, also combined with modern machine learning tools [72], will enable transition from highly controlled laboratory observations to in-situ field measurements, providing a more accurate and nuanced understanding of particle behavior under natural conditions. By integrating these advanced sensing and modeling capabilities with the proposed theoretical frameworks, the approaches introduced herein could be used to revolutionize the ways by which aeolian (and fluvial) sediment transport can be predicted and managed, in a diverse range of environments.

### 5.5. Application of energy criterion for the aeolian transport of plastics

The pervasive distribution of plastics in the environment has become a significant ecological and human health challenge, of increasing importance to address at the source e.g., before these anthropogenic debris are transported downwind. The role of aeolian processes in the transport of plastics, particularly meso- to micro- plastics, is an emerging field of study that requires comprehensive modelling to understand and mitigate its impacts. The presented study, although focusing on the modelling of transport processes for natural coarse particles, also highlights the importance of wind tunnel experiments in assessing the aerodynamic threshold for the incipient motion of plastic debris.

In the context of plastic debris, obtaining a better understanding of the threshold wind energy required for transporting plastic particles downwind, is essential for predicting how plastics can be mobilised by wind and subsequently dispersed across different landscapes. The incipient motion of plastic debris is influenced by several factors, including particle size, shape, density, and surface roughness, as well as environmental conditions such as wind speed and turbulence. Even though herein the density and size of the plastic sphere has not been changed, it very closely resembles the density and characteristic dimensions of typical plastic debris carried by the wind, which has been shown to be significant depending both on the environmental context and the abundance and features of the plastic debris [73-76]. Thus, the results reported herein can be directly useful for those aiming to model the transport and deposition of micro-plastics carried within the eroded soil as well as macro-plastics transported under various environmental settings (see Figure 13) [77, 78].

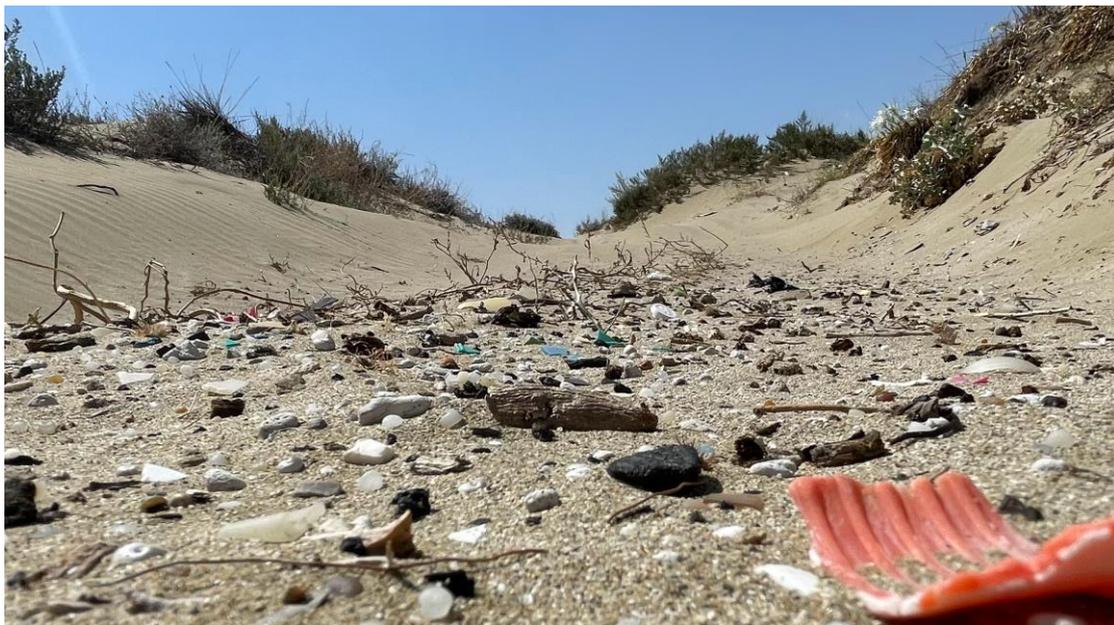

**Figure 13.** Demonstration of the prevalence of windblown light plastic debris (note the plastic translucent beads and parts of plastic cups), at a sandy region near a coastal area. Photo taken from the coastline looking upwind.

Furthermore, the dynamic criterion of energy is a novel approach that accounts for the effects of turbulent fluctuations on particle motion. Turbulence in the atmosphere can significantly affect the transport of particles, as it can enhance lift and lead to more erratic and long-range transport patterns. By incorporating the scaling of turbulent fluctuations with particle dimensions, researchers can create more accurate models that reflect the complexities of plastic transport in the natural environment. Thus, these experiments can be seen as a critical step in developing a fundamental understanding of the dynamics involved for the transport of plastics. Through the comprehensive modelling of these transport processes hotspots of plastic accumulation can be identified, the pathways through which plastics travel be better understood, and the effectiveness of interventions aimed at reducing plastic pollution evaluated.

Even though more wind tunnel experiments are needed to assess the effects of particle shape, density, and size more broadly on these processes, are needed and planned for the near future, this study underscores the application of the dynamic criterion of energy in modelling the aeolian transport of plastics. This research can be seen as a first foundational step for developing predictive models that can inform environmental policy and management practices. As the challenge of plastic pollution continues to grow, it is imperative to even further advance our understanding of how plastics are transported by wind and to leverage this knowledge to protect nature's ecological integrity and human well-being.

## 6. Conclusions

Creep or rolling are among the major transport modes for the aerodynamic entrainment of coarse grains resting on a flat granular surface near incipient flow conditions. Such movement is an important process for the evolution of many aeolian bedforms. Traditional time-averaged force models, however, are deficient in characterizing the intermittent incipient motions in turbulent airflow. To address this issue, this study develops an event-based energy model and corresponding energy criterion to define the incipient entrainment of coarse particles by rolling. It is hypothesized that full entrainment by rolling is caused by energetic events in turbulent airflow, which can impart enough energy to sediment particles exposed on the soil surface. To validate the criterion, the study conducted a novel set of experiments, recording the time series of local wind velocities and considering the energy balance for particle entrainment. Through the experiments, the validity of the concept is demonstrated, and a critical rolling is obtained, allowing the empirical threshold energy level to be determined. Furthermore, for appropriate values of the energy transfer coefficient, the theoretically predicted threshold curves accurately predict whether a flow event will lead to incomplete entrainment (rocking) or full entrainment (rolling). It is concluded that there is a critical aerodynamic force for rocking to occur, but minimum energy offered from energetic flow events is required for full entrainment by rolling. Rocking particles, even though they cannot cause a net transport downwind, can transfer energy from the airflow to the granular surface by means of mechanical sieving, which is responsible for the formation of gravel-mantled megaripples. Though the experiments involve idealized grain properties and their configuration, this does not limit the generality of this study's major findings and implications. Further experimental work, however, is needed to establish values for the energy transfer coefficient for a wider range of flows and configurations corresponding to natural environments.

Thus, the proposed framework is a novel approach that considers the role of turbulent, energetic airflow structures for:

1. indirectly evaluating incipient motion conditions of natural coarse particles (full entrainment downwind)

2. directly assessing incipient motion conditions of light micro- to meso- sized plastics

3. indirectly identify the conditions responsible for setting the particles into rocking motions (incomplete entrainments), leading to the processes of creep transport or mechanical sieving.

It is also found that the creep transport efficiency is about $C_{eff}$ ~0.0005, an order smaller compared to rolling transport efficiency $C_{eff}$ ~0.005. Transport efficiency for rocking ($C_{eff}$ ~0.001) is about 1/5$^{th}$ of the rolling threshold and an order of magnitude less than the saltation threshold efficiency ($C_{eff}$ ~0.01).

As such, this research can be seen as a step for developing more comprehensive predictive models for aeolian transport of plastics and sediment based on first principles, that can better inform policy and management practices, as well as decision making for engineering researchers and practitioners.

**Acknowledgements**. This research is supported by the National Natural Science Foundation of China (Grants No. 41171005, 41071005, 12272344, 12350710176) and the Ministry of Science and Technology of the People's Republic of China (Grant No. 2013CB956000).

**Data availability statement:** Data (Valyrakis and Xiao-Hu, 2024 [79]) used in this manuscript, as well as for the wind velocity profile and turbulence spectra are archived in a repository and can be accessed via the following URLs: https://zenodo.org/records/7782901.